\documentclass[10pt]{article}

\usepackage{amsmath}
\usepackage{physics}
\usepackage{amsfonts}
\usepackage{amssymb}
\usepackage{graphicx}%
\usepackage{amstext}
\usepackage{latexsym}
\usepackage{color}
\usepackage{mathtools}
\usepackage{graphics}
\usepackage{cite}
\usepackage{slashed}
\usepackage{epsfig}
\usepackage{amsthm}
\usepackage{appendix}
\usepackage{hyperref}
\usepackage{multirow}
\usepackage{color}
\usepackage[normal]{subfigure}
\usepackage{rotating}
\usepackage[table]{xcolor}
\usepackage{enumitem}
\usepackage[utf8x]{inputenc}
\usepackage{authblk}
\usepackage{colortbl}
\usepackage{pdflscape}
\usepackage{bm}
\usepackage{gensymb}

\usepackage[english]{babel}
\allowdisplaybreaks

\newcommand{\be}{\begin{equation}}
\newcommand{\en}{\end{equation}}

\newtheorem{defi}{Definition}[section]
\newtheorem{lem}[defi]{Lemma}

\newtheorem{Theo}{Theorem}[section]

\newtheorem{remark}[Theo]{Remark}

\newcommand{\bedefin}{\begin{defi}}
\newcommand{\findefi}{\end{defi} \medskip}

\newcommand{\belem}{\begin{lem}$\!\!${\bf }}
\newcommand{\enlem}{\end{lem}}

\newcommand{\beno}{\begin{equation*}}
\newcommand{\enno}{\end{equation*}}

\newcommand{\bea}{\begin{eqnarray}}
\newcommand{\ena}{\end{eqnarray}}

\catcode `\@=11 \@addtoreset{equation}{section}

\catcode `\@=12

\newcommand{\g}{G_{\hbox{\tiny{NC}}}}

\newcommand{\G}{\mathfrak{g}_{\hbox{\tiny{NC}}}}

\begin{document}

\title{On a charged spinless point particle minimally coupled to a constant magnetic field in a noncommutative plane}

\author[1]{S. Hasibul Hassan Chowdhury\thanks{shhchowdhury@bracu.ac.bd}}
\author[2,3,4]{Talal Ahmed Chowdhury\thanks{talal@du.ac.bd}}

\affil[1]{Department of Mathematics and Natural Sciences, BRAC University, 66 Mohakhali, Dhaka 1212, Bangladesh}
\affil[2]{Department of Physics, University of Dhaka, Dhaka 1000, Bangladesh}
\affil[3]{Department of Physics and Astronomy, University of Kansas, Lawrence, Kansas 66045, USA}
\affil[4]{The Abdus Salam International Centre for Theoretical Physics, Strada Costiera 11, I-34151 Trieste, Italy}

%\date{\today}

\maketitle

\begin{abstract}
In this paper, we provide a mathematically and physically consistent minimal prescription for a charged spinless point particle coupled to a constant magnetic field in a 2-dimensional noncommutative plane. It turns out to be a gauge invariant prescription in contrast to the widely and carelessly used naive minimal prescription in the context of 2-dimensional quantum mechanics in a noncommutative plane. Besides, we explore the noncommutative U(1) gauge theoretic structure of the underlying noncommutative system by explicitly computing the 1-parameter family of Seiberg-Witten maps.
\end{abstract}

\section{Introduction}\label{sec:intro}
A quantum phase space (see \cite{alvarezetal}) is specified by an operator algebra. The operator algebra that concerns us is generated by 4 self-adjoint operators $\hat{X}^{r}$, $\hat{Y}^{r}$, $\hat{p}_{x}$, $\hat{p}_{y}$ and the identity operator $\hat{\mathbb{I}}$ on $L^{2}(\mathbb{R}^{2},dx\;dy)$ obeying the following set of commutation relations:
\begin{equation}\label{equation:noncommutative-plane-comm-rel}
[\hat{X}^{r},\hat{Y}^{r}]=i\vartheta\hat{\mathbb{I}},\hspace{.5in}[\hat{X}^{r},\hat{p}_{x}]=[\hat{Y}^{r},\hat{p}_{y}]=i\hbar\hat{\mathbb{I}},
\end{equation}
the other commutators vanish identically. The pertaining operator algebra represents a family of noncommutative 2-planes indexed by the real parameter $r$. The second set of equalities in (\ref{equation:noncommutative-plane-comm-rel}) refers to the quantum mechanical structure in the underlying operator algebra while the first equality spells out the noncommutative structure of the corresponding planes. Hence, equation (\ref{equation:noncommutative-plane-comm-rel}) points out that the operator algebra that we are dealing with studies quantum mechanics in noncommutative spaces. Here, the real parameter $r$ does not affect the underlying commutation relations and hence is called the {\em gauge parameter}. Equation (\ref{equation:noncommutative-plane-comm-rel}) yields a 1-parameter family of irreducible self-adjoint representations of the universal enveloping algebra $\mathcal{U}(\G)$ of the Lie algebra $\G$ of the 7-dimensional real nilpotent Lie group $\g$ studied extensively in earlier papers \cite{Chowdhury1, Chowdhury2, Chowdhury3, Chowdhury4}. All the representations in the family are unitarily equivalent to each other. This family of irreducible representations is labelled by the ordered triple $(\hbar,\vartheta,0)$. In this sense, all the noncommutative 2-planes in the family are equivalent or more precisely gauge equivalent. Observe that the magnetic field pertaining to the noncommutativity of momenta coordinates is taken to be zero in this case. This is the reason the last coordinate of the unitary dual of $\g$ is taken to be zero in this case, i.e., $(\hbar,\vartheta,0)$. One such representative of the family due to $r=1$ is worked out in case III.3 of \cite{Chowdhury2}. Compare the irreducible representation of $\g$ appearing in case III.3 with the 1-parameter family of noncommutative planes in (\ref{equation:g_nc_rep_nc_plane}) by plugging in $r=1$. 

Next, we minimally couple this 1 parameter ($r$) family of noncommutative planes with an external uniform \textit{magnetic field} $B$ using noncommutative U(1) gauge fields (parametrized also by $r$) whose operatorial representation is given by 
\begin{equation}\label{equation:gauge-field-express}
\hat{\mathbf{A}}^{r}=\left(\frac{-2(1-r)\hbar B}{\hbar+\sqrt{\hbar^{2}-4r(r-1)e\hbar\vartheta B}}\hat{Y}^{r},\frac{2r\hbar B}{\hbar+\sqrt{\hbar^{2}-4r(r-1)e\hbar\vartheta B}}\hat{X}^{r}\right).
\end{equation}
The familiar {\em Landau} and {\em symmetric gauges} correspond to the gauge parameter values $r=1$ and $r=\frac{1}{2}$, respectively. We emphasise at this stage that the unitary dual of $\g$ corresponding to the 1-parameter ($r$) family of noncommutative planes that concerns us is labeled by $(\hbar,\vartheta,0)$. The magnetic field $B$ appearing in the operatorial expression (\ref{equation:gauge-field-express}) of 1-parameter family of gauge fields is an external one, i.e., it has nothing to do with intrinsic phase space noncommutativity. Here, $e$ is the coupling parameter and we follow minimal coupling prescription , i.e., the respective covariant derivative operators are written as
\begin{equation}\label{equation:covariant-derivative-operator}
\hat{\Pi}^{r}_{i}=\hat{p}_{i}-e\hat{\mathbf{A}}^{r}_{i},
\end{equation} 
where $i=x,y$. And the covariant derivative operators obey the following commutation relation:
\begin{equation}\label{equation:covariant-derivative-rel}
[\hat{\Pi}^{r}_{x},\hat{\Pi}^{r}_{y}]=i\hbar B\hat{\mathbb{I}}.
\end{equation}
Again, the gauge parameter $r$ disappears from the expression of the commutator between the covariant derivative operators along the $x$ and $y$ directions. It is important to note that 
\begin{equation}\label{equation:dont-obey-CCR}
[\hat{X}^{r},\hat{\Pi}^{r}_{x}]\neq i\hbar\hat{\mathbb{I}},\hspace{.5in}[\hat{Y}^{r},\hat{\Pi}^{r}_{y}]\neq i\hbar\hat{\mathbb{I}},
\end{equation}
as is reflected in the last $2$ equations of (\ref{equation:modiefied-commut-rel}). It is perfectly fine as $\hat{\Pi}^{r}_{x}$ and $\hat{\Pi}^{r}_{y}$ obtained using (\ref{equation:covariant-derivative-operator}) do not represent generators of the Lie group $\g$ while $\hat{p}_{x}$ and $\hat{p}_{y}$ indeed do and satisfy the governing commutation relations (\ref{equation:noncommutative-plane-comm-rel}) of 1-parameter family of noncommutative planes. Gauge independence of the commutator in (\ref{equation:covariant-derivative-rel}) ensures the gauge invariance of the energy spectra of the Landau Hamiltonian as is proven both using operator formalism and star-product formalism in section (\ref{sec:star-prod}). There are instances in the literature \cite{Dayietal,dulat,dulat2009quantum, gangopadhyay,ncanistropicoscltr} where the authors write the operator representing the gauge field as
\begin{equation}\label{equation:wrong-minimal prescription}
\hat{\mathbf{A}}^{r}=((r-1)B\hat{Y}^{r},rB\hat{X}^{r}),
\end{equation}
with the self-adjoint operators $\hat{X}^{r}$ and $\hat{Y}^{r}$ are as in (\ref{equation:g_nc_rep_nc_plane}). Substitution of $\hat{\mathbf{A}}^{r}$ in (\ref{equation:covariant-derivative-operator}) leads to
\begin{equation}\label{equation:gauge-dependence-commutator}
[\hat{\Pi}^{r}_{x},\hat{\Pi}^{r}_{y}]=ie\hbar B\left[1-\frac{er(r-1)\vartheta B}{\hbar}\right]\hat{\mathbb{I}}.
\end{equation}
In contrast to (\ref{equation:covariant-derivative-rel}), presence of the gauge parameter $r$ in (\ref{equation:gauge-dependence-commutator}) contributes gauge dependence to the energy spectra of the Landau Hamiltonian. Such minimal prescription is termed as \textit{naive minimal prescription} in \cite{Chowdhury5}. 

Landau problem in a noncommutative plane can be looked upon as a $U(1)_{\star}$ gauge theory. In fact, the relevant gauge field can be treated as an ordered pair of formal power series in $\vartheta$ with coefficients in $C^{\infty}(\mathbb{R}^{2})$. The set of such formal power series can be endowed with the structure of a ring and is denoted by $C^{\infty}(\mathbb{R}^{2})[[\vartheta]]$. Using the deformation quantization technique, we deform the classical algebra $C^{\infty}(\mathbb{R}^{2})$ of smooth functions on the configuration space $\mathbb{R}^{2}$ by introducing a 1-parameter family of noncommutative associative star-products $*^{r}$ (see \ref{equation:star-prod-gen-def}) parametrized by the same gauge parameter $r$. The power series representation of (\ref{equation:gauge-field-express}) is given by
\begin{equation}\label{equation:gauge-field-function-form}
\mathbf{A}^{\hbox{\tiny{nc}}}\equiv(\mathbf{A}^{\hbox{\tiny{nc}}}_{x},\mathbf{A}^{\hbox{\tiny{nc}}}_{y})=\left(\frac{-2(1-r)\hbar B}{\hbar+\sqrt{\hbar^{2}-4r(r-1)e\hbar\vartheta B}}y,\frac{2r\hbar B}{\hbar+\sqrt{\hbar^{2}-4r(r-1)e\hbar\vartheta B}}x\right).
\end{equation}
Here, both the components of the ordered pair above belong to $C^{\infty}(\mathbb{R}^{2})[[\vartheta]]$, i.e., each component above can be expressed as a formal power series in $\vartheta$ with coefficients in $C^{\infty}(\mathbb{R}^{2})$. The governing star-commutation relations of this noncommutative ring of formal power series are all provided in (\ref{star-representation},\ref{equation:momentum-functions},\ref{equation:star-commutation}). All these star-commutation relations between the relevant observables are in complete agreement with the operatorial commutation relations introduced in (\ref{equation:noncommutative-plane-comm-rel}). Let us take the liberty of sloppily calling a formal power series a function. Whenever we refer to a function in $(C^{\infty}(\mathbb{R}^{2}),*^{r})$, we essentially refer to a formal power series in $\vartheta$ with coefficients lying in $C^{\infty}(\mathbb{R}^{2})$ with the product in the noncommutative ring given by $*^{r}$. At the expense of being imprecise, we will benefit ourselves from obtaining the flavour of the deformation of the commutative algebra (with respect to the pointwise product) of classical observables in $C^{\infty}(\mathbb{R}^{2})$ by the introduction of the 1-parameter family of star-products $*^{r}$ introduced in (\ref{equation:star-prod-gen-def}).

We obtain a function $\lambda^{\hbox{\tiny{nc}}}$ in $C^{\infty}(\mathbb{R}^{2})$ (strictly speaking a formal power series in $\vartheta$ with coefficients in $C^{\infty}(\mathbb{R}^{2})$) that produces the desired $U(1)_{\star}$ phase factor $U=e^{i\frac{\lambda^{\hbox{\tiny{nc}}}}{\hbar}}_{\star}$ of the underlying noncommutative $U(1)$ gauge theoretic structure of the noncommutative Landau problem. Here, $e^{x}_{\star}=\mathbb{I}+x+\frac{1}{2!}x*^{r}x+...$, with $\mathbb{I}$ being the constant map in $C^{\infty}(\mathbb{R}^{2})$ that maps everything to the constant real number $1$. We call this function {\em noncommutative gauge function} which is given by the following expression:
\begin{equation}\label{nc-gauge-func}
\lambda^{\hbox{\tiny{nc}}}=\lambda+3e\epsilon r(r-1)\vartheta\frac{B^{2} xy}{\hbar}+\mathcal{O}(\vartheta^{2}),
\end{equation}
associated with an infinitesimal change $\epsilon=r^{\prime}-r$ of the gauge parameter. Here, $\lambda=\epsilon Bxy$ is called the {\em undeformed gauge function}. It is worth remarking at this stage that in the literature of noncommutative gauge theory (see, for example, \cite{Seiberg-Witten, Delduc}), $\lambda^{\hbox{\tiny{nc}}}$ is referred to as gauge parameter. In contrast to the existing literature, our gauge parameter $r$ is the one that parametrizes not only the members of the same equivalence class $(\hbar,\vartheta,0)$ (representing the noncommutative 2-planes) of the unitary dual of $\g$ (see \ref{equation:g_nc_rep_nc_plane}) but also the noncommutative associative star-products (see \ref{equation:star-prod-gen-def}) $*^{r}$ that the deformed function space $(C^{\infty}(\mathbb{R}^{2}),*^{r})$ over the noncommutative 2-planes are equipped with. The same gauge parameter $r$ parametrizes an invertible family of Seiberg-Witten like maps (see \cite{Seiberg-Witten}) between the associated noncommutative gauge field, field strength and their commutative counterparts that we explore in detail in section \ref{sec:Seiberg-Witten-map}.

The first purpose of the present paper is to give a consistent prescription for a noncommutative plane to be minimally coupled with an external magnetic field $B$\footnote{For earlier studies, please see \cite{Duval:2000xr, Duval:2001hu, Horvathy:2002wc}}. By consistency we mean the gauge invariance of the energy spectra of the underlying Landau Hamiltonian. This mathematical and physical consistency is discussed using both operator and star-product formalism. The next goal of the paper is to address some unsettled issues in \cite{Mezincescu, Delduc}. The computations were carried out there in symmetric gauge ($r=\frac{1}{2}$) and it was inferred there that similar results should hold in other gauges also. For example, in \cite{Mezincescu} within the symmetric gauge configuration, it was shown that although the particle's effective mass $m$ and the coupling $e$ change, the ratio $\frac{e^{2}B^{2}}{m}$ remains unchanged. We verified this result for any value of the gauge parameter $r$. The final goal of this paper is to find a family of Seiberg-Witten maps \cite{Seiberg-Witten} parametrized by $r\in\mathbb{R}$ relating the underlying noncommutative gauge fields and magnetic field strength with their commutative counterparts. We verify that the Moyal star-product \cite{Moyal, Groenewold} case (corresponding to $r=\frac{1}{2}$) reproduces the well-known Seiberg-Witten map.

The paper is organized as follows. In section (\ref{sec:minimal}), we briefly discuss two ways of approaching the problem of a charged spinless point particle of mass $m$ placed in a noncommutative plane subjected to a constant magnetic field $B$ (refer to section 6 of the excellent review article \cite{Delduc}). The first approach (section 6.1 of \cite{Delduc}) involving a nonminimal prescription has been addressed in \cite{Chowdhury5}. Here, the uniform magnetic field $B$ refers to the one originated from phase space noncommutativity and comes from the unitary dual of the underlying kinematical symmetry group $\g$. The second approach involves minimally coupling the particle of mass $m$ to an \textit{external uniform magnetic field} $B$. This external magnetic field $B$ is not related to the unitary dual of $\g$. In section (\ref{sec:star-prod}), we discuss how one can obtain c-equivalent (in the sense of \cite{Bayen-Flato1,Bayen-Flato2,Bayen-Flato3}) star products $*^{r}$ parametrized by the gauge parameter $r$. The symmetric gauge version ($r=\frac{1}{2}$) of the star products is the familiar Moyal product. Later in this section, we show that the spectra of the Landau Hamiltonian is independent of the gauge parameter $r$. Along the way, we prove that the ratio $\frac{e^{2}B^{2}}{m}$  remains gauge invariant, i.e., independent of $r$ where $e$ is the coupling through which the spinless point particle of mass $m$ is coupled to the external magnetic field $B$. In section (\ref{sec:Seiberg-Witten-map}), we obtain a family of invertible maps parametrized by $r$ that relates the noncommutative U(1) gauge fields and the relevant field strength with their commutative counterparts. Again, $r=\frac{1}{2}$, here, yields the familiar Seiberg-Witten map. Finally in section (\ref{sec:conclusion}), we provide concluding remarks along with comments on possible future research directions.

\section{Minimal prescription for quantum mechanics in noncommutative space}\label{sec:minimal}

There are two equivalent ways of tackling the quantum mechanical problem of a charged spinless point particle placed in a noncommutative plane under the influence of a perpendicular constant magnetic field $B$ (\cite{Delduc}). In one approach, one has the following set of commutation relations to start with
\begin{equation}\label{equation:NCQM-first-approach}
\begin{split}
[\hat{X}^{s},\hat{\Pi}^{r,s}_{x}]=[\hat{Y}^{s},\hat{\Pi}^{r,s}_{y}]=i\hbar\hat{\mathbb{I}},\;\;[\hat{\Pi}^{r,s}_{x},\hat{\Pi}^{r,s}_{y}]=i\hbar B\hat{\mathbb{I}},\;\;[\hat{X}^{s},\hat{Y}^{s}]=i\vartheta\hat{\mathbb{I}}.
\end{split}
\end{equation}
Here, $\hat{X}^{s}$, $\hat{Y}^{s}$, $\hat{\Pi}^{r,s}_{x}$ and $\hat{\Pi}^{r,s}_{y}$ are some families of self-adjoint unbounded operators on $L^{2}(\mathbb{R}^{2},dx\;dy)$ representing the non-central generators of a nilpotent Lie group denoted by $\g$ (\cite{Chowdhury2}). These families of operators are indexed by $2$ parameters $r$ and $s$ in such a way (see equation (2.8) of \cite{Chowdhury5}) that they disappear from the right side of the commutators given in (\ref{equation:NCQM-first-approach}). This is the reason the 2 parameters $r,s$ were called gauge parameters in \cite{Chowdhury5}. Certain values of the ordered pair $(r,s)\in\mathbb{R}^{2}$, there, were seen to produce the symmetric and Landau gauge representations of $\g$. Also, $\hat{\mathbb{I}}$ is the identity operator on $L^{2}(\mathbb{R}^{2},dx\;dy)$. In this approach, the magnetic field $B$ appears as one of the 3 parameters ($\hbar,\vartheta,B$) labeling the unitary dual of $\g$. This approach has been studied extensively in \cite{Chowdhury5}. The 2 parameter ($r,s$) family of irreducible representations of $\g$ in terms of the self-adjoint operators $\hat{X}^{s}$, $\hat{Y}^{s}$, $\hat{\Pi}_{x}^{r,s}$, $\hat{\Pi}_{y}^{r,s}$ satisfying the commutation relations \ref{equation:NCQM-first-approach} are easily seen to be gauge equivalent, i.e., operators corresponding to distinct ordered pairs $(r,s)\in\mathbb{R}^{2}$ and $(r^{\prime},s^{\prime})\in\mathbb{R}^{2}$ are \textit{unitarily equivalent}.

In the other approach, one starts with a noncommutative plane characterized by the following one parameter family of equivalent irreducible representations of $\G$, the Lie algebra of $\g$:
\begin{equation}\label{equation:g_nc_rep_nc_plane}
\begin{split}
\hat{X}^{r}&=\hat{x}+\frac{(r-1)\vartheta}{\hbar}\hat{p}_{y},\\
\hat{Y}^{r}&=\hat{y}+\frac{r\vartheta}{\hbar}\hat{p}_{x},\\
\hat{\Pi}_{x}&=\hat{p}_{x},\\
\hat{\Pi}_{y}&=\hat{p}_{y}.
\end{split}
\end{equation}

The ordered triple $(\hbar,\vartheta,0)\in\mathbb{R}^{2}$ in the unitary dual of $\g$ determines the underlying equivalence class of irreducible unitary representations of $\g$. Here, the quantum mechanical position operators $\hat{x}$ and $\hat{y}$ act on a generic element $\psi\in L^{2}(\mathbb{R}^{2},dx\;dy)$ in the following way:
\begin{equation}\label{equation:qm-position}
\begin{split}
(\hat{x}\psi)(x,y)&=x\psi(x,y),\\
(\hat{y}\psi)(x,y)&=y\psi(x,y),
\end{split}
\end{equation}
while the quantum mechanical momenta operators $\hat{p}_{x}$ and $\hat{p}_{y}$ act as derivations on a generic element $\psi\in L^{2}(\mathbb{R}^{2},dx\;dy)$ in the following way:
\begin{equation}\label{equation:qm-position}
\begin{split}
(\hat{p}_{x}\psi)(x,y)&=-i\hbar\frac{\partial\psi}{\partial x}(x,y),\\
(\hat{p}_{y}\psi)(x,y)&=-i\hbar\frac{\partial\psi}{\partial y}(x,y).
\end{split}
\end{equation}

The representation of $\mathcal{U}(\G)$ given in (\ref{equation:g_nc_rep_nc_plane}) obeys the following set of commutation relations: 
\begin{equation}
\begin{split}
&[\hat{X}^{r},\hat{Y}^{r}]=i\vartheta\hat{\mathbb{I}}\;,\;[\hat{X}^{r},\hat{\Pi}_{x}]=i\hbar\hat{\mathbb{I}},\\
&[\hat{Y}^{r},\hat{\Pi}_{y}]=i\hbar\hat{\mathbb{I}}\;,\;[\hat{\Pi}_{x},\hat{\Pi}_{y}]=0.
\end{split}
\end{equation} 

Now, one couples the nonrelativistic spinless point particle of mass $m$ situated in the noncommutative plane given by (\ref{equation:g_nc_rep_nc_plane}) to a constant magnetic field $B$ minimally ($e$ being the coupling) through the following 1-parameter family of representations:
\begin{equation}\label{equation:minimal-coupling-rep}
\begin{split}
&\hat{X}^{r}=\hat{x}+\frac{(r-1)\vartheta}{\hbar}\hat{p}_{y},\\
&\hat{Y}^{r}=\hat{y}+\frac{r\vartheta}{\hbar}\hat{p}_{x},\\
&\hat{\Pi}^{r}_{x}=\frac{2(1-r)e\hbar B}{\hbar+\sqrt{\hbar^{2}-4r(r-1)e\hbar\vartheta B}}\hat{y}+\left[1+\frac{2r(1-r)e\vartheta B}{\hbar+\sqrt{\hbar^{2}-4r(r-1)e\hbar\vartheta B}}\right]\hat{p}_{x},\\
&\hat{\Pi}^{r}_{y}=\frac{-2re\hbar B}{\hbar+\sqrt{\hbar^{2}-4r(r-1)e\hbar\vartheta B}}\hat{x}+\left[1+\frac{2r(1-r)e\vartheta B}{\hbar+\sqrt{\hbar^{2}-4r(r-1)e\hbar\vartheta B}}\right]\hat{p}_{y}.
\end{split}
\end{equation}

Now the minimal coupling of the noncommutative plane given by (\ref{equation:g_nc_rep_nc_plane}) to the electromagnetic field can be seen by a simple manipulation of (\ref{equation:minimal-coupling-rep}):
\begin{equation}\label{equation:manipulation-min-coupl-rep}
\begin{split}
&\hat{X}^{r}=\hat{x}+\frac{(r-1)\vartheta}{\hbar}\hat{p}_{y},\\
&\hat{Y}^{r}=\hat{y}+\frac{r\vartheta}{\hbar}\hat{p}_{x},\\
&\hat{\Pi}^{r}_{x}=\hat{p}_{x}+\frac{2e(1-r)\hbar B}{\hbar+\sqrt{\hbar^{2}-4r(r-1)e\hbar\vartheta B}}\left(\hat{y}+\frac{r\vartheta}{\hbar}\hat{p}_{x}\right),\\
&\hat{\Pi}^{r}_{y}=\hat{p}_{y}-\frac{2re\hbar B}{\hbar+\sqrt{\hbar^{2}-4r(r-1)e\hbar\vartheta B}}\left[\hat{x}+\frac{(r-1)\vartheta}{\hbar}\hat{p}_{y}\right].
\end{split}
\end{equation}

The representation (\ref{equation:manipulation-min-coupl-rep}) satisfy the following set of commutation relations:
\begin{equation}\label{equation:modiefied-commut-rel}
\begin{split}
&[\hat{X}^{r},\hat{Y}^{r}]=i\vartheta\hat{\mathbb{I}},\hspace{.15in}[\hat{\Pi}^{r}_{x},\hat{\Pi}^{r}_{y}]=ie\hbar B\hat{\mathbb{I}},\\
&[\hat{X}^{r},\hat{\Pi}_{x}^{r}]=\left[1+\frac{2(1-r)e\vartheta B}{\hbar+\sqrt{\hbar^{2}-4r(r-1)e\hbar\vartheta B}}\right]i\hbar\hat{\mathbb{I}},\\
&[\hat{Y}^{r},\hat{\Pi}_{y}^{r}]=\left[1+\frac{2re\vartheta B}{\hbar+\sqrt{\hbar^{2}-4r(r-1)e\hbar\vartheta B}}\right]i\hbar\hat{\mathbb{I}}.
\end{split}
\end{equation}

Writing $\hat{\mathbf{\Pi}}^{r}\equiv(\hat{\Pi}^{r}_{x},\hat{\Pi}^{r}_{y})$ and $\hat{\mathbf{p}}\equiv(\hat{p}_{x},\hat{p}_{y})$, one obtains from the expressions of kinematical momenta provided by 
(\ref{equation:manipulation-min-coupl-rep}) that
\begin{equation}\label{equation: kinematical-momenta}
\hat{\mathbf{\Pi}}^{r}=\hat{\mathbf{p}}-e\hat{\mathbf{A}}^{r},
\end{equation}
where the noncommutative U(1) gauge fields are given by
\begin{equation}\label{equation:gauge-field-def}
\hat{\mathbf{A}}^{r}=\left(\frac{-2(1-r)\hbar B}{\hbar+\sqrt{\hbar^{2}-4r(r-1)e\hbar\vartheta B}}\hat{Y}^{r},\frac{2r\hbar B}{\hbar+\sqrt{\hbar^{2}-4r(r-1)e\hbar\vartheta B}}\hat{X}^{r}\right).
\end{equation}
The familiar symmetric and Landau gauges correspond to the following values of $r$:
\begin{equation*}
r_{\hbox{\tiny{sym}}}=\frac{1}{2},\;\hbox{and}\; r_{\hbox{\tiny{Lan}}}=1.
\end{equation*}
The gauge field expressions can be read off using (\ref{equation:gauge-field-def}) as
\begin{equation}\label{equation:Landau-symmetric-gauge-field}
\begin{split}
&\hat{\mathbf{A}}^{\frac{1}{2}}=\left(-\frac{\hbar B}{\hbar+\sqrt{\hbar^{2}+e\hbar\vartheta B}}\hat{Y}^{\frac{1}{2}},\frac{\hbar B}{\hbar+\sqrt{\hbar^{2}+e\hbar\vartheta B}}\hat{X}^{\frac{1}{2}}\right),\\
&\hat{\mathbf{A}}^{1}=(0,B\hat{x}).
\end{split}
\end{equation}
The representation (\ref{equation:minimal-coupling-rep}) takes the following form in Landau gauge:
\begin{equation}\label{equation:Landau-gauge-rep}
\begin{split}
\hat{X}^{1}=\hat{x},\;\;\hat{Y}^{1}=\hat{y}+\frac{\vartheta}{\hbar}\hat{p}_{x},\;\;\hat{\Pi}_{x}^{1}=\hat{p}_{x},\;\;\hat{\Pi}_{y}^{2}=-eB\hat{x}+\hat{p}_{y},
\end{split}
\end{equation}
while in the symmetric gauge (\ref{equation:minimal-coupling-rep}) looks as follow:
\begin{equation}
\begin{split}
&\hat{X}^{\frac{1}{2}}=\hat{x}-\frac{\vartheta}{2\hbar}\hat{p}_{y},\\
&\hat{Y}^{\frac{1}{2}}=\hat{y}+\frac{\vartheta}{2\hbar}\hat{p}_{x},\\
&\hat{\Pi}_{x}^{\frac{1}{2}}=\frac{e\hbar B}{\hbar+\sqrt{\hbar^{2}+e\hbar\vartheta B}}\hat{y}+\left[1+\frac{e\vartheta B}{2(\hbar+\sqrt{\hbar^{2}+e\hbar\vartheta B})}\right]\hat{p}_{x},\\
&\hat{\Pi}_{y}^{\frac{1}{2}}=-\frac{e\hbar B}{\hbar+\sqrt{\hbar^{2}+e\hbar\vartheta B}}\hat{x}+\left[1+\frac{e\vartheta B}{2(\hbar+\sqrt{\hbar^{2}+e\hbar\vartheta B})}\right]\hat{p}_{y}.
\end{split}
\end{equation}

\begin{remark}
It is important to note that the 1-parameter family of gauge fields provided by (\ref{equation:gauge-field-def}) satisfies

\begin{equation}\label{gauge-field-consistency-eqn}
\partial_{x}\hat{\mathbf{A}}^{r}_{y}-\partial_{y}\hat{\mathbf{A}}^{r}_{x}-\frac{ie}{\hbar}[\hat{\mathbf{A}}^{r}_{x},\hat{\mathbf{A}}^{r}_{y}]=B\hat{\mathbb{I}}=:\hat{F}_{xy},
\end{equation}

where $\partial_x=\frac{\partial}{\partial x}$ and $\partial_y=\frac{\partial}{\partial y}$ and $\hat{\mathbf{A}}^{r}_{x}$ and $\hat{\mathbf{A}}^{r}_{y}$ are the components of the noncommutative vector potential $\hat{\mathbf{A}}^{r}$ defined in (\ref{equation:gauge-field-def}). One also defines $\partial_{i}\hat{\mathbf{A}}^{r}_{j}=[\partial_{i},\hat{\mathbf{A}}^{r}_{j}]$ for $i,j=x,y$. Here, the $c$-number operator $\hat{F}_{xy}$ stands for the noncommutative field strength associated with the constant magnetic field $B$.
\end{remark}

\section{Star product approach to quantum mechanics on a noncommutative space}\label{sec:star-prod}
We consider the function space to be $C^{\infty}(\mathbb{R}^{2})$. Now, we define $4$ functions $x, y, \Pi_{x},\Pi_{y}\in C^{\infty}(\mathbb{R}^{2})$ by means of each of their noncommutative $*$ product with some other (sufficiently well-behaved) vector $\psi\in L^{2}(\mathbb{R}^{2},dx\;dy)$ as follow:
\begin{equation}\label{star-representation}
\begin{split}
&x*^{r}\psi:=\hat{X}^{r}\psi=x\psi-i(r-1)\vartheta\frac{\partial\psi}{\partial y},\\
&y*^{r}\psi:=\hat{Y}^{r}\psi=y\psi-ir\vartheta\frac{\partial\psi}{\partial x},\\
&\Pi_{x}*^{r}\psi:=\hat{\Pi}^{r}_{x}\psi=\frac{2(1-r)e\hbar B}{\hbar+\sqrt{\hbar^{2}-4r(r-1)e\hbar\vartheta B}}y\psi-i\hbar\left[1+\frac{2r(1-r)e\vartheta B}{\hbar+\sqrt{\hbar^{2}-4r(r-1)e\hbar\vartheta B}}\right]\frac{\partial\psi}{\partial x},\\
&\Pi_{y}*^{r}\psi:=\hat{\Pi}^{r}_{y}\psi=\frac{-2re\hbar B}{\hbar+\sqrt{\hbar^{2}-4r(r-1)e\hbar\vartheta B}}x\psi-i\hbar\left[1+\frac{2r(1-r)e\vartheta B}{\hbar+\sqrt{\hbar^{2}-4r(r-1)e\hbar\vartheta B}}\right]\frac{\partial\psi}{\partial y}.
\end{split}
\end{equation}

The last two equations in (\ref{star-representation}) merit elaboration: the $2$ functions $\Pi_{x}$ and $\Pi_{y}$ are actually derived objects using the relations $\Pi_{x}=p_{x}-e\mathbf{A}^{\hbox{\tiny{nc}}}_{x}$ and $\Pi_{y}=p_{y}-e\mathbf{A}^{\hbox{\tiny{nc}}}_{y}$. The 2 smooth functions $p_x$ and $p_y$ are given by the following star-products with sufficiently well-behaved wave function $\psi\in L^{2}(\mathbb{R}^{2},dx\;dy)$:
\begin{equation}\label{equation:momentum-functions}
p_{x}*\psi=\hat{p}_{x}\psi=-i\hbar\frac{\partial\psi}{\partial x},\;\;\;\;p_{y}*\psi=\hat{p}_{y}\psi=-i\hbar\frac{\partial\psi}{\partial y}.
\end{equation}
And, the vector valued function $\mathbf{A}^{\hbox{\tiny{nc}}}$ is given by
\begin{equation}\label{equation:gauge-field-function}
\mathbf{A}^{\hbox{\tiny{nc}}}\equiv(\mathbf{A}^{\hbox{\tiny{nc}}}_{x},\mathbf{A}^{\hbox{\tiny{nc}}}_{y})=\left(\frac{-2(1-r)\hbar B}{\hbar+\sqrt{\hbar^{2}-4r(r-1)e\hbar\vartheta B}}y,\frac{2r\hbar B}{\hbar+\sqrt{\hbar^{2}-4r(r-1)e\hbar\vartheta B}}x\right).
\end{equation}

Using (\ref{equation:momentum-functions}), (\ref{equation:gauge-field-function}), (\ref{star-representation}) and linearity of star product, one now easily verifies that the following hold:
\begin{equation}\label{equation:covariant-derivative-star-prod}
\begin{aligned}
&\Pi_{x}*^{r}\psi=(p_{x}-e\mathbf{A}^{\hbox{\tiny{nc}}}_{x})*^{r}\psi=p_{x}*\psi-e\mathbf{A}^{\hbox{\tiny{nc}}}_{x}*^{r}\psi,\\
&\Pi_{y}*^{r}\psi=(p_{y}-e\mathbf{A}^{\hbox{\tiny{nc}}}_{y})*^{r}\psi=p_{y}*\psi-e\mathbf{A}^{\hbox{\tiny{nc}}}_{y}*^{r}\psi.
\end{aligned}
\end{equation}

By virtue of the commutation relations (\ref{equation:modiefied-commut-rel}), the nonvanishing star-commutation relations can be read off as
\begin{equation}\label{equation:star-commutation}
\begin{split}
&[x\!\overset{\;\;*^{r}}{,}y]=i\vartheta\mathbb{I},\;\;[\Pi_{x}\!\!\overset{\;\;*^{r}}{,}\Pi_{y}]=ie\hbar B\mathbb{I},\\
&[x\!\overset{\;\;*^{r}}{,}\Pi_{x}]=\left[1+\frac{2(1-r)e\vartheta B}{\hbar+\sqrt{\hbar^{2}-4r(r-1)e\hbar\vartheta B}}\right]i\hbar\mathbb{I}=i\hbar\mathbb{I}-e[x\!\overset{\;\;*^{r}}{,}\!\mathbf{A}^{\hbox{\tiny{nc}}}_{x}],\\
&[y\!\overset{\;\;*^{r}}{,}\Pi_{y}]=\left[1+\frac{2re\vartheta B}{\hbar+\sqrt{\hbar^{2}-4r(r-1)e\hbar\vartheta B}}\right]i\hbar\mathbb{I}=i\hbar\mathbb{I}-e[y\!\overset{\;\;*^{r}}{,}\!\mathbf{A}^{\hbox{\tiny{nc}}}_{y}].
\end{split}
\end{equation}

Also, $\mathbb{I}$ is the constant map in $C^{\infty}(\mathbb{R}^{2})$ that maps everything to the constant value $1$ and the star-commutator between 2 functions $F(x,y)$ and $G(x,y)$ in $C^{\infty}(\mathbb{R}^{2})$ is defined as

\begin{equation}\label{equation:star-commut-def}
 [F(x,y)\!\overset{\;\;*^{r}}{,}G(x,y)]=F(x,y)*^{r}\!G(x,y)-G(x,y)*^{r}\!F(x,y),
 \end{equation}
 where the gauge parameter $r$ dependent star-product between $F(x,y),G(x,y)\in C^{\infty}(\mathbb{R}^{2})$ is given by
 \begin{equation}\label{equation:star-prod-gen-def}
 F(x,y)*^{r}\!G(x,y)=F(x,y)e^{-i(r-1)\vartheta\overset{\leftarrow}{\partial_{x}}\overset{\rightarrow}{\partial_{y}}-ir\vartheta\overset{\leftarrow}{\partial_{y}}\overset{\rightarrow}{\partial_{x}}}G(x,y).
\end{equation}

In fact, (\ref{equation:star-prod-gen-def}) tells us that $*:C^{\infty}(\mathbb{R}^{2})\times C^{\infty}(\mathbb{R}^{2})\rightarrow C^{\infty}(\mathbb{R}^{2})[[\vartheta]]$, where $C^{\infty}(\mathbb{R}^{2})[[\vartheta]]$ is the ring of all formal power series in the deformation parameter $\vartheta$ with coefficients in the ring $C^{\infty}(\mathbb{R}^{2})$. In fact, $*$ can be extended to the ring $C^{\infty}(\mathbb{R}^{2})[[\vartheta]]$ so that it becomes a map $*:C^{\infty}(\mathbb{R}^{2})[[\vartheta]]\times C^{\infty}(\mathbb{R}^{2})[[\vartheta]]\rightarrow C^{\infty}(\mathbb{R}^{2})[[\vartheta]]$ \cite{Bayen-Flato1}. One can now go on to show that $2$ star products $*^{r}$ and $*^{r^{\prime}}$ given by ($\ref{equation:star-prod-gen-def}$) are gauge equivalent (or C-equivalent in the sense of \cite{Bayen-Flato1}), i.e., there exists an invertible operator $T$ on $C^{\infty}(\mathbb{R}^{2})$ such that
\begin{equation}\label{equation:c-equivalent-star-product}
T(F*^{r}G)=T(F)*^{r^{\prime}}T(G).
\end{equation}
Let us now observe that $T=e^{i(r-r^{\prime})\vartheta\overset{\rightarrow}{\partial_{x}}\overset{\rightarrow}{\partial_{y}}}$ is the required invertible operator which satisfies (\ref{equation:c-equivalent-star-product}).
Indeed,  one has
\begin{equation}\label{equation:derivation-T-operator-left}
\begin{aligned}[b]
T(F*^{r}G)&=e^{i(r-r^{\prime})\vartheta\overset{\rightarrow}{\partial_{x}}\overset{\rightarrow}{\partial_{y}}}[Fe^{-i(r-1)\vartheta\overset{\leftarrow}{\partial_{x}}\overset{\rightarrow}{\partial_{y}}-ir\vartheta\overset{\leftarrow}{\partial_{y}}\overset{\rightarrow}{\partial_{x}}}G]\\
&=Fe^{i(r-r^{\prime})\vartheta\overset{\leftarrow}{\partial_{x}}\overset{\leftarrow}{\partial_{y}}+i(r-r^{\prime})\vartheta\overset{\leftarrow}{\partial_{x}}\overset{\rightarrow}{\partial_{y}}+i(r-r^{\prime})\vartheta\overset{\leftarrow}{\partial_{y}}\overset{\rightarrow}{\partial_{x}}+i(r-r^{\prime})\vartheta\overset{\rightarrow}{\partial_{x}}\overset{\rightarrow}{\partial_{y}}}e^{-i(r-1)\vartheta\overset{\leftarrow}{\partial_{x}}\overset{\rightarrow}{\partial_{y}}-ir\vartheta\overset{\leftarrow}{\partial_{y}}\overset{\rightarrow}{\partial_{x}}}G\\
&=Fe^{i(r-r^{\prime})\vartheta\overset{\leftarrow}{\partial_{x}}\overset{\leftarrow}{\partial_{y}}+i(r-r^{\prime})\vartheta\overset{\leftarrow}{\partial_{x}}\overset{\rightarrow}{\partial_{y}}+i(r-r^{\prime})\vartheta\overset{\leftarrow}{\partial_{y}}\overset{\rightarrow}{\partial_{x}}+i(r-r^{\prime})\vartheta\overset{\rightarrow}{\partial_{x}}\overset{\rightarrow}{\partial_{y}}-i(r-1)\vartheta\overset{\leftarrow}{\partial_{x}}\overset{\rightarrow}{\partial_{y}}-ir\vartheta\overset{\leftarrow}{\partial_{y}}\overset{\rightarrow}{\partial_{x}}}G\\
&=Fe^{i(r-r^{\prime})\vartheta\overset{\leftarrow}{\partial_{x}}\overset{\leftarrow}{\partial_{y}}-i(r^{\prime}-1)\vartheta\overset{\leftarrow}{\partial_{x}}\overset{\rightarrow}{\partial_{y}}-ir^{\prime}\vartheta\overset{\leftarrow}{\partial_{y}}\overset{\rightarrow}{\partial_{x}}+i(r-r^{\prime})\vartheta\overset{\rightarrow}{\partial_{x}}\overset{\rightarrow}{\partial_{y}}}G.
\end{aligned}
\end{equation}

On the other hand,
\begin{equation}\label{equation:derivation-T-operator-right}
\begin{aligned}[b]
T(F)*^{r^{\prime}}T(G)&=[e^{i(r-r^{\prime})\vartheta\overset{\rightarrow}{\partial_{x}}\overset{\rightarrow}{\partial_{y}}}F]e^{i(r^{\prime}-1)\vartheta\overset{\leftarrow}{\partial_{x}}\overset{\rightarrow}{\partial_{y}}-ir^{\prime}\vartheta\overset{\leftarrow}{\partial_{y}}\overset{\rightarrow}{\partial_{x}}}[e^{i(r-r^{\prime})\vartheta\overset{\rightarrow}{\partial_{x}}\overset{\rightarrow}{\partial_{y}}}G]\\
&=Fe^{i(r-r^{\prime})\vartheta\overset{\leftarrow}{\partial_{x}}\overset{\leftarrow}{\partial_{y}}-i(r^{\prime}-1)\vartheta\overset{\leftarrow}{\partial_{x}}\overset{\rightarrow}{\partial_{y}}-ir^{\prime}\vartheta\overset{\leftarrow}{\partial_{y}}\overset{\rightarrow}{\partial_{x}}+i(r-r^{\prime})\vartheta\overset{\rightarrow}{\partial_{x}}\overset{\rightarrow}{\partial_{y}}}G.
\end{aligned}
\end{equation}
From (\ref{equation:derivation-T-operator-left}) and (\ref{equation:derivation-T-operator-right}) follows 
(\ref{equation:c-equivalent-star-product}).

\begin{remark}
It is important to note that one can reproduce the familiar Moyal product by taking $r=\frac{1}{2}$ in the general definition of 1-parameter family of star-products given by (\ref{equation:star-prod-gen-def}). In other words, Moyal star-product is the symmetric gauge version of the star product (\ref{equation:star-prod-gen-def}).
\end{remark}

The associativity of the star product follows from the general definition of it presented in (\ref{equation:star-prod-gen-def}). Now let us consider the Hamiltonian for the 2-dimensional Landau problem: a charged spinless point particle of mass $m$ and charge $e$ placed in a constant magnetic field. Using star product the Landau Hamiltonian is the following smooth function on $\mathbb{R}^{2}$:
\begin{equation}\label{equation:Landau-Hamiltonian}
H=\frac{1}{2m}(p_{x}-e\mathbf{A}^{\hbox{\tiny{nc}}}_{x})*^{r}(p_{x}-e\mathbf{A}^{\hbox{\tiny{nc}}}_{x})+\frac{1}{2m}(p_{y}-e\mathbf{A}^{\hbox{\tiny{nc}}}_{y})*^{r}(p_{y}-e\mathbf{A}^{\hbox{\tiny{nc}}}_{y}).
\end{equation} 
Hence, the pertinent Schrodinger equation for a sufficiently well-behaved function\\ $\psi\in L^{2}(\mathbb{R}^{2},dx\;dy)$ reads
\begin{equation}\label{equation:Schrodinger-eqn-Landau-problem}
H*^{r}\psi=E\psi.
\end{equation}
Let us first see that
\begin{equation}\label{equation:reduced-mass-charge-derivation}
\begin{split}
(p_{x}-e\mathbf{A}^{\hbox{\tiny{nc}}}_{x})*^{r}\psi&=(\hat{p}_{x}-e\hat{\mathbf{A}}^{r}_{x})\psi\\
&=\left[\hat{p}_{x}+\frac{2e(1-r)\hbar B}{\hbar+\sqrt{\hbar^{2}-4r(r-1)e\hbar\vartheta B}}\hat{Y}^{r}\right]\psi\\
&=\left[1+\frac{2r(1-r)e\vartheta B}{\hbar+\sqrt{\hbar^{2}-4r(r-1)e\hbar\vartheta B}}\right]\left[\hat{p}_{x}-\frac{e}{1+\frac{2r(1-r)e\vartheta B}{\hbar+\sqrt{\hbar^{2}-4r(r-1)e\hbar\vartheta B}}}\mathbf{A}^{\hbox{\tiny{nc}}}_{x}\right]\psi.
\end{split}
\end{equation}
Let us denote
\begin{equation}\label{equation:factor}
\bar{\Lambda}(r)=1+\frac{2r(1-r)e\vartheta B}{\hbar+\sqrt{\hbar^{2}-4r(r-1)e\hbar\vartheta B}}.
\end{equation}
Then (\ref{equation:reduced-mass-charge-derivation}) reads
\begin{equation}\label{equation:expression-reduced-mass-charge-intermediate}
(p_{x}-e\mathbf{A}^{\hbox{\tiny{nc}}}_{x})*^{r}\psi=\bar{\Lambda}(r)\left(\hat{p}_{x}-\frac{e}{\bar{\Lambda}(r)}\mathbf{A}^{\hbox{\tiny{nc}}}_{x}\right)\psi,
\end{equation}
which leads to the following

\begin{equation}\label{equation:Schrodinger-summand-1}
\begin{split}
\frac{1}{2m}(p_{x}-e\mathbf{A}^{\hbox{\tiny{nc}}}_{x})*^{r}(p_{x}-e\mathbf{A}^{\hbox{\tiny{nc}}}_{x})*^{r}\psi&=\frac{[\bar{\Lambda}(r)]^{2}}{2m}\left(\hat{p}_{x}-\frac{e}{\bar{\Lambda}(r)}\mathbf{A}^{\hbox{\tiny{nc}}}_{x}\right)^{2}\psi,\\
&=\frac{1}{2m_{*}}(\hat{p}_{x}-e_{*}\mathbf{A}^{\hbox{\tiny{nc}}}_{x})^{2}\psi,
\end{split}
\end{equation}
where the reduced mass and reduced charge, denoted by $m_{*}$ and $e_{*}$, respectively, are given by
\begin{equation}\label{equation:reduced-mass-charge-def}
m_{*}=\frac{m}{[\bar{\Lambda}(r)]^{2}}\hspace{.5in}e_{*}=\frac{e}{\bar{\Lambda}(r)}.
\end{equation}
One can similarly go on to find 
\begin{equation}\label{equation:Schrodinger-summand-2}
\frac{1}{2m}(p_{y}-e\mathbf{A}^{\hbox{\tiny{nc}}}_{y})*^{r}(p_{y}-e\mathbf{A}^{\hbox{\tiny{nc}}}_{y})*^{r}\psi=\frac{1}{2m_{*}}(\hat{p}_{y}-e_{*}\mathbf{A}^{\hbox{\tiny{nc}}}_{y})^{2}\psi,
\end{equation}
where $m_{*}$ and $e_{*}$ are as given by (\ref{equation:reduced-mass-charge-def}).

One finally arrives at the following:
\begin{equation}\label{equation:nc-Landau-problem-deformed-Hamiltonian}
\begin{split}
&\frac{1}{2m}(p_{x}-e\mathbf{A}^{\hbox{\tiny{nc}}}_{x})*^{r}(p_{x}-e\mathbf{A}^{\hbox{\tiny{nc}}}_{x})*^{r}\psi+\frac{1}{2m}(p_{y}-e\mathbf{A}^{\hbox{\tiny{nc}}}_{y})*^{r}(p_{y}-e\mathbf{A}^{\hbox{\tiny{nc}}}_{y})*^{r}\psi\\
&=\frac{1}{2m_{*}}[(\hat{p}_{x}-e_{*}\mathbf{A}^{\hbox{\tiny{nc}}}_{x})^{2}+(\hat{p}_{y}-e_{*}\mathbf{A}^{\hbox{\tiny{nc}}}_{y})^{2}]\psi=:\hat{H}^{r}_{*}\psi.
\end{split}
\end{equation}
Here, $\hat{H}^{r}_{*}$ with the expressions for the reduced charge $e_{*}$ and mass $m_{*}$ as given in (\ref{equation:reduced-mass-charge-def}) is called the deformed Hamiltonian of the noncommutative Landau problem.

Now, one finds out from (\ref{equation:gauge-field-function}) that
\begin{equation}\label{equation:nc-magntc-field}
\partial_{x}\mathbf{A}^{\hbox{\tiny{nc}}}_{y}-\partial_{y}\mathbf{A}^{\hbox{\tiny{nc}}}_{x}=\frac{2\hbar B}{\hbar+\sqrt{\hbar^{2}-4r(r-1)e\hbar\vartheta B}}\mathbb{I}=:\bar{B}(r,\vartheta)\mathbb{I}.
\end{equation}

Gathering (\ref{equation:nc-magntc-field}) with (\ref{equation:gauge-field-function}) and the first equation of (\ref{equation:star-commutation}), one verifies that the star product version of (\ref{gauge-field-consistency-eqn}) given by the following holds

\begin{equation}\label{equation:non-abelian-yang-mills-star-version}
\partial_{x}\mathbf{A}^{\hbox{\tiny{nc}}}_{y}-\partial_{y}\mathbf{A}^{\hbox{\tiny{nc}}}_{x}-\frac{ie}{\hbar}[\mathbf{A}^{\hbox{\tiny{nc}}}_{x}\!\!\overset{\;\;*^{r}}{,}\!\mathbf{A}^{\hbox{\tiny{nc}}}_{y}]=B\mathbb{I}.
\end{equation}

Using the 1-dimensional harmonic oscillator technique of solving the canonical Landau problem (page 21 in \cite{Delduc}), one arrives at the following eigenvalues of the deformed Hamiltonian $H_{*}$ as given in (\ref{equation:nc-Landau-problem-deformed-Hamiltonian}) (we choose the speed of light $c=1$):
\begin{equation}\label{equation:def-hamiltonian-eigenvalues}
E_{n}=\hbar\frac{|e_{*}\bar{B}(r,\vartheta)|}{m_{*}}\left(n+\frac{1}{2}\right)\hspace{.5in}\hbox{with}\;\; n=0,1,2,...
\end{equation}
By using (\ref{equation:factor}), (\ref{equation:reduced-mass-charge-def}) and (\ref{equation:nc-magntc-field}), one finds that the eigenvalues given in (\ref{equation:def-hamiltonian-eigenvalues}) are indeed gauge invariant, i.e., independent of the gauge parameter $r$:
\begin{equation}\label{equation:gauge-invariant-eigenvalues}
\begin{split}
E_{n}&=\hbar\frac{|e\bar{\Lambda}(r)\bar{B}(r,\vartheta)|}{m}\left(n+\frac{1}{2}\right),\\
&=\hbar\frac{|eB|}{m}\left(n+\frac{1}{2}\right).
\end{split}
\end{equation}
Each of the eigenstates associated with the eigenvalues above are infinitely degenerate as expected. The eigenvalues are also found to be independent of the noncommutativity parameter $\vartheta$.

Using (\ref{equation:nc-magntc-field}) along with (\ref{equation:factor}), one finds that
\begin{equation}\label{equation:important-nc-relation}
\begin{aligned}[b]
\bar{\Lambda}(r)\bar{B}(r,\vartheta)&=\frac{2\hbar B[\hbar-2r(r-1)e\vartheta B+\sqrt{\hbar^{2}-4r(r-1)e\hbar\vartheta B}]}{[\hbar+\sqrt{\hbar^{2}-4r(r-1)e\hbar\vartheta B}]^{2}},\\
&=\frac{2\hbar B[\hbar-2r(r-1)e\vartheta B+\sqrt{\hbar^{2}-4r(r-1)e\hbar\vartheta B}]}{[2\hbar^{2}-4r(r-1)e\hbar\vartheta B+2\hbar\sqrt{\hbar^{2}-4r(r-1)e\hbar\vartheta B]}},\\
&=B.
\end{aligned}
\end{equation}

(\ref{equation:important-nc-relation}) is shown to hold true \cite{Mezincescu,Delduc} in the symmetric gauge only, i.e., for $r=\frac{1}{2}$. Also, if we manipulate the expression for $\bar{B}(r,\vartheta)$ (see (\ref{equation:nc-magntc-field})), one finds out 
\begin{equation}\label{equation:agreement-with-Delduc}
B=\left[1-\frac{r(r-1)e\vartheta\bar{B}}{\hbar}\right]\bar{B},
\end{equation}
so that combining (\ref{equation:agreement-with-Delduc}) with (\ref{equation:important-nc-relation}), one obtains an expression for $\bar{\Lambda}(r)$ using the effective magnetic field $\bar{B}(r,\vartheta)$ in a given gauge:
\begin{equation}\label{equation:factor-using-b-bar}
\bar{\Lambda}(r)=1-\frac{r(r-1)e\vartheta\bar{B}}{\hbar},
\end{equation}
the symmetric gauge value of which agrees with \cite{Mezincescu,Delduc}:
\begin{equation}\label{equation:symmetric-gauge-factor}
\bar{\Lambda}\left(\frac{1}{2}\right)=1+\frac{e\vartheta\bar{B}}{4\hbar}.
\end{equation}

Assuming $r=\frac{1}{2}$, on the other hand, in the expression for $\bar{B}(r,\vartheta)$ in (\ref{equation:nc-magntc-field}), one obtains
\begin{equation}\label{equation:symmetric-gauge-effective-field}
\begin{split}
\bar{B}\left(\frac{1}{2},\vartheta\right)&=\frac{2\hbar B}{\hbar+\sqrt{\hbar^{2}+e\hbar\vartheta B}},\\
&=\frac{2}{e\vartheta}(\sqrt{\hbar^{2}+e\hbar\vartheta B}-\hbar),
\end{split}
\end{equation}
which again agrees with the effective magnetic field expression for the symmetric gauge given in \cite{Mezincescu, Delduc,Scholtz-dual}.

Also, the term $\frac{e^{2}B^{2}}{m}$ is inferred to be gauge invariant in \cite{Mezincescu}. Indeed, we find here that
\begin{equation}\label{equation:gauge-invariant-term-verified}
\frac{e_{*}^{2}B^{2}}{m_{*}}=\frac{\frac{e^{2}}{[\bar{\Lambda}(r)]^{2}}B^{2}}{\frac{m}{[\bar{\Lambda}(r)]^{2}}}=\frac{e^{2}B^{2}}{m}.
\end{equation}

Although we have seen explicitly that the spectra of the Landau Hamiltonian in the context of noncommutative space turn out to be gauge invariant by means of (\ref{equation:gauge-invariant-eigenvalues}), let us try to investigate this gauge invariance group theoretically and study its relation with the star product approach introduced earlier in this section.

From the earlier discussion, we have seen that there exists an invertible operator $T$ on $C^{\infty}(\mathbb{R}^{2})$ that satisfies
\begin{equation}\label{equation:operator-formalism-star-prod-formalism}
T(H)*^{r^{\prime}}T(\psi)=T(H*^{r}\psi)=ET(\psi),
\end{equation}
where $H$ is the Hamiltonian function (\ref{equation:Landau-Hamiltonian}) in $C^{\infty}(\mathbb{R}^{2})$ for the noncommutative Landau problem and $\psi\in L^{2}(\mathbb{R}^{2},dx\;dy)$ is a sufficiently well-behaved wavefunction. If one writes $T(H)=\widetilde{H}$ and $T(\psi)=\widetilde{\psi}$, then (\ref{equation:operator-formalism-star-prod-formalism}) reads
\begin{equation}\label{equation:covariant-S-equation}
\widetilde{H}*^{r^{\prime}}\widetilde{\psi}=E\widetilde{\psi}.
\end{equation}
In view of (\ref{equation:Schrodinger-eqn-Landau-problem}) and (\ref{equation:covariant-S-equation}), one finds that under gauge transformation Schrodinger equation transforms covariantly so that the underlying spectra $E$ remain invariant. 

Let us now denote by $\hat{H}_{*}^{r}$ and $\hat{\widetilde{H}}_{*}^{r^{\prime}}$, the operators representing the smooth functions $H$ and $\widetilde{H}$ in $C^{\infty}(\mathbb{R}^{2})$, respectively. One can refer to (\ref{equation:nc-Landau-problem-deformed-Hamiltonian}) for an explicit expression of this operator. In order to obtain $\hat{\widetilde{H}}_{*}^{r^{\prime}}$, one just replaces $r$ with $r^{\prime}$ throughout in (\ref{equation:nc-Landau-problem-deformed-Hamiltonian}). The one-parameter family of Hamiltonians $\hat{H}^{r}$ is expressed using the one parameter family of equivalent irreducible self-adjoint representations of $\G$ due to a fixed ordered pair $(\hbar,\vartheta)$ (\ref{equation:g_nc_rep_nc_plane}). Physically all these gauge equivalent representations (parametrized by $r$) represent the same noncommutative plane. Hence the unbounded self-adjoint operator $\hat{H}_{*}^{r}$ and $\hat{\widetilde{H}}_{*}^{r^{\prime}}$ are related by
\begin{equation}\label{equation:intertwining}
\hat{\widetilde{H}}^{r^{\prime}}_{*}=\hat{U}\hat{H}_{*}^{r}\hat{U}^{-1},
\end{equation}
with $\hat{U}$ being a unitary operator on $L^{2}(\mathbb{R}^{2},dx\;dy)$. In operatorial language, (\ref{equation:Schrodinger-eqn-Landau-problem}) translates to
\begin{equation}\label{S-eq-op-form}
\hat{H}_{*}^{r}\psi=E\psi,
\end{equation}
leading to 
\begin{equation}\label{equation:intermediate-eq}
\hat{U}\hat{H}_{*}^{r}\psi=E\hat{U}\psi,
\end{equation}
so that one obtains
\begin{equation*}
\hat{U}\hat{H}_{*}^{r}\hat{U}^{-1}\hat{U}\psi=E\hat{U}\psi.
\end{equation*}
In other words, the eigenvalue of the gauge transformed Hamiltonian $\hat{\widetilde{H}}^{r^{\prime}}$ remains the same, i.e., $E$ corresponding to the eigenvector $\widetilde{\psi}=\hat{U}\psi$: 
\begin{equation}\label{equation:gauge-equiv-S-eq}
\hat{\widetilde{H}}^{r^{\prime}}\!\!\!\widetilde{\psi}=E\widetilde{\psi}.
\end{equation}
The star-product version of (\ref{equation:gauge-equiv-S-eq}) is of course (\ref{equation:covariant-S-equation}). If $U$ is an invertiable smooth function in $C^{\infty}(\mathbb{R}^{2})$ that is represented by the unitary operator $\hat{U}$ appearing in (\ref{equation:intertwining}), then the star-product version of (\ref{equation:intertwining}) reads
\begin{equation}\label{equation:star-product-version-intertwining}
T(H)=\widetilde{H}=U*^{r}H*^{r}U^{-1}.
\end{equation}
At the wavefunction level, one also has
\begin{equation}\label{equation:wavefunction-star-version}
T(\psi)=\widetilde{\psi}=U*^{r}\psi,
\end{equation}
so that in terms of the invertible smooth function $U\in C^{\infty}(\mathbb{R}^{2})$, (\ref{equation:covariant-S-equation})now reads
\begin{equation}\label{equation:S-eq-using-U-func}
U*^{r}H*^{r}U^{-1}*^{r^{\prime}}U*^{r}\psi=E(U*^{r}\psi).
\end{equation}
Since $U^{-1}*^{r^{\prime}}U=\mathbb{I}$, the map in $C^{\infty}(\mathbb{R}^{2})$ that maps everything to 1, one finds that the following holds
\begin{equation}\label{equation:star-version-S-eq-using-U-func}
U*^{r}H*^{r}\psi=E(U*^{r}\psi),
\end{equation}
which is indeed the star-product version of (\ref{equation:intermediate-eq}).

\section{On 1-parameter family of Seiberg-Witten maps for noncommutative U(1) gauge theory}\label{sec:Seiberg-Witten-map}

N. Seiberg and E. Witten were the first to provide a set of formulae in \cite{Seiberg-Witten} establishing the relationship between the noncommutative quantities (gauge fields, field strength etc.) and their commutative counterparts. A variant of the magnetic field $B$ is used to this end in \cite{Delduc} (see equation (65) at p. 20) which we call commutative Seiberg-Witten field strength denoted by $\mathfrak{B}$. This gauge dependent quantity in terms of the physical magnetic field $B$ reads
\begin{equation}\label{equation:Sei-Witt-mag-field}
\mathfrak{B}=\frac{\hbar B}{\hbar-4r(r-1)e\vartheta B}.
\end{equation}

Note that the symmetric gauge (corresponding to $r=\frac{1}{2}$) version of (\ref{equation:Sei-Witt-mag-field}) matches with the one provided in \cite{Delduc} (equation (65) at p.20). In terms of $\mathfrak{B}$, the commutative Seiberg-Witten gauge fields are defined as (in \cite{Delduc} expressed in symmetric gauge, i.e., for $r=\frac{1}{2}$)
\begin{equation}\label{equation:redefined-commut-Sei-Witt-gauge-fields}
\mathbf{A}^{r}\equiv(\mathbf{A}_{x},\mathbf{A}_{y})=((r-1)\mathfrak{B}y,r\mathfrak{B}x).
\end{equation}
We will be using $\mathbf{A}$ instead of $\mathbf{A}^{r}$ for the sake of notational clarity to denote the redefined commutative gauge field, i.e., the 1-parameter family of commutative Seiberg-Witten gauge fields.

The 1-parameter family of noncommutative Seiberg-Witten gauge fields are precisely the same as the ones introduced in (\ref{equation:gauge-field-function}):
\begin{equation}\label{equation:nc-Seiberg-Witt-gauge-field}
\mathbf{A}^{\hbox{\tiny{nc}}}\equiv(\mathbf{A}^{\hbox{\tiny{nc}}}_{x},\mathbf{A}^{\hbox{\tiny{nc}}}_{y})=((r-1)\bar{B}y,r\bar{B}x),
\end{equation}
where $\bar{B}$ in terms of the commutative Seiberg-Witten field strength $\mathfrak{B}$ is given by:
 \begin{equation}\label{equation:def-commut-Seib-Witt-mag-field}
\bar{B}(r,\vartheta)=\frac{\hbar}{2r(1-r)e\vartheta}\left[\frac{1}{\sqrt{1+\frac{4r(r-1)e\vartheta\mathfrak{B}}{\hbar}}}-1\right],
\end{equation}

The {\em noncommutative gauge function} $\lambda^{\hbox{\tiny{nc}}}$ can then be obtained using the infinitesimal variation $\epsilon=r^{\prime}-r$ of the gauge parameter associated with the noncommutative gauge fields given in (\ref{equation:gauge-field-function-form} or \ref{equation:nc-Seiberg-Witt-gauge-field}):
\begin{equation}\label{equation:def-nc-gauge-function}
\delta\mathbf{A}^{\hbox{\tiny{nc}}}_{j}=\partial_{j}\lambda^{\hbox{\tiny{nc}}}+\frac{ie}{\hbar}[\lambda^{\hbox{\tiny{nc}}}\!\overset{\;\;*^{r}}{,}\mathbf{A}^{\hbox{\tiny{nc}}}_{j}],\\
\end{equation}
where $j=x,y$. A straightforward but lengthy computation then leads to the following expression relating the noncommutative gauge function with its commutative counterpart:
\begin{equation}\label{equation:rel-com-nc-gauge-func}
\lambda^{\hbox{\tiny{nc}}}=\lambda+3e\epsilon r(r-1)\vartheta\frac{B^{2} xy}{\hbar}+\mathcal{O}(\vartheta^{2}),
\end{equation}
where the {\em undeformed gauge function} $\lambda$ is defined as $\lambda=\epsilon Bxy$.
The detailed derivation of (\ref{equation:rel-com-nc-gauge-func}) is provided in the Appendix. Now, a sufficiently well-behaved wavefunction $\psi\in C^{\infty}(\mathbb{R}^{2})$ transforms under the underlying $U(1)_{\star}$ gauge transformation as
\begin{equation}\label{equation:wavefunction-gauge-transfo}
\psi\rightarrow U_{\lambda^{\hbox{\tiny{nc}}}}*^{r}\psi,
\end{equation}
where $U_{\lambda^{\hbox{\tiny{nc}}}}=e^{\frac{ie}{\hbar}\lambda^{\hbox{\tiny{nc}}}}_{\star}$. Here, $e^{x}_{\star}$ denotes star-exponential, i.e., in the power series expansion of $e^{x}$, all the products are taken to be star-products. Using (\ref{equation:def-nc-gauge-function}), one verifies that the star-exponent $\frac{e\lambda^{\hbox{\tiny{nc}}}}{\hbar}$ is indeed dimensionless. Under $U(1)_{\star}$ gauge transformation the noncommutative gauge fields transform as
\begin{equation}\label{equation:finite-gauge-field-transfo}
\mathbf{A}_{j}^{\hbox{\tiny{nc}}}\rightarrow U_{\lambda^{\hbox{\tiny{nc}}}}*^{r}\mathbf{A}^{\hbox{\tiny{nc}}}_{j}*^{r}U_{\lambda^{\hbox{\tiny{nc}}}}^{-1}+\frac{i\hbar}{e}U_{\lambda^{\hbox{\tiny{nc}}}}*^{r}\partial_{j}U^{-1}_{\lambda^{\hbox{\tiny{nc}}}},
\end{equation}
with $j=x,y$. Indeed, one verifies that (\ref{equation:finite-gauge-field-transfo}) is in agreement with the infinitesimal gauge transformation given in (\ref{equation:def-nc-gauge-function}) due to an infinitesimal variation $\epsilon=r^{\prime}-r$ of the gauge parameter. Also, note that the inverse of $U_{\lambda^{\hbox{\tiny{nc}}}}$ is given by $U_{\lambda^{\hbox{\tiny{nc}}}}^{-1}=e^{-\frac{ie}{\hbar}\lambda^{\hbox{\tiny{nc}}}}_{\star}$.

Of course, (\ref{equation:nc-Seiberg-Witt-gauge-field}) satisfies,
\begin{equation}\label{equation:nc-Sei-Witt-gauge-field-comm}
[p_{x}-e\mathbf{A}^{\hbox{\tiny{nc}}}_{x}\!\overset{\;\;*^{r}}{,}p_{y}-e\mathbf{A}^{\hbox{\tiny{nc}}}_{y}]=ie\hbar B\mathbb{I}.
\end{equation}

Now, one has
\begin{equation}\label{equation:first-comp-gauge-field}
\begin{aligned}[b]
\mathbf{A}^{\hbox{\tiny{nc}}}_{x}&=(r-1)\frac{\hbar}{2r(1-r)e\vartheta}\left[\frac{1}{\sqrt{1+\frac{4r(r-1)e\vartheta\mathfrak{B}}{\hbar}}}-1\right]y,\\
&=\frac{\hbar}{2re\vartheta}\left[1-\frac{1}{\sqrt{1+\frac{4r(r-1)e\vartheta\mathfrak{B}}{\hbar}}}\right]y,\\
&=\frac{\hbar}{2er\vartheta}\left[1-\left\{1+\frac{4r(r-1)e\vartheta\mathfrak{B}}{\hbar}\right\}^{-\frac{1}{2}}\right]y,\\
&=\frac{\hbar y}{2er\vartheta}\left[\frac{2r(r-1)e\vartheta\mathfrak{B}}{\hbar}-\frac{6r^{2}(r-1)^{2}e^{2}\vartheta^{2}\mathfrak{B}^{2}}{\hbar^{2}}+\cdots\right],\\
&=(r-1)\mathfrak{B}y-\frac{3r(r-1)^{2}e\vartheta\mathfrak{B}^{2}y}{\hbar}+\mathcal{O}(\vartheta^{2}),\\
&=\mathbf{A}_{x}-\frac{2er}{\hbar}\theta^{xy}\mathbf{A}_{x}\partial_{y}\mathbf{A}_{x}-\frac{e(r-1)}{\hbar}\theta^{xy}\mathbf{A}_{x}\partial_{x}\mathbf{A}_{y}+\mathcal{O}(\vartheta^{2}),\\
&=\mathbf{A}_{x}-\frac{e}{\hbar}\theta^{xy}\mathbf{A}_{x}\left[(3r-1)\partial_{y}\mathbf{A}_{x}+(1-r)F_{yx}\right]+\mathcal{O}(\vartheta^{2}),
\end{aligned}
\end{equation}
where $\theta^{xy}=-\theta^{yx}=\vartheta$, are the entries of the real $2\times 2$ antisymmetric matrix $\theta$ given by
\begin{equation}\label{equation:def-antisymm-mat}
\theta=\begin{bmatrix}0&\vartheta\\-\vartheta&0\end{bmatrix},
\end{equation}
and 
\begin{equation}\label{equation:entries-commut-S-W-field-strength}
F_{xy}=\partial_{x}\mathbf{A}_{y}-\partial_{y}\mathbf{A}_{x}=\mathfrak{B}\mathbb{I}=-F_{yx}, 
\end{equation}
 are the entries of the following $2\times 2$ antisymmetric real matrix $F$ representing the commutative Seiberg-Witten field strength:
\begin{equation}\label{equation:Field-strength-tensor}
F=\begin{bmatrix}0&\partial_{x}\mathbf{A}_{y}-\partial_{y}\mathbf{A}_{x}\\\partial_{y}\mathbf{A}_{x}-\partial_{x}\mathbf{A}_{y}&0\end{bmatrix}.
\end{equation}

Also, from (\ref{equation:nc-Seiberg-Witt-gauge-field}) and (\ref{equation:def-commut-Seib-Witt-mag-field}), it follows that
\begingroup
\begin{equation}\label{equation:y-component-S-W-map-gauge-field}
\allowdisplaybreaks
\begin{aligned}[b]
\mathbf{A}^{\hbox{\tiny{nc}}}_{y}&=r\frac{\hbar}{2r(1-r)e\vartheta}\left[\frac{1}{\sqrt{1+\frac{4r(r-1)e\vartheta\mathfrak{B}}{\hbar}}}-1\right]x,\\
&=\frac{\hbar}{2(1-r)e\vartheta}\left[\left\{1+\frac{4r(r-1)e\vartheta\mathfrak{B}}{\hbar}\right\}^{-\frac{1}{2}}-1\right]x,\\
&=\frac{\hbar x}{2(1-r)e\vartheta}\left[-\frac{2r(r-1)e\vartheta\mathfrak{B}}{\hbar}+\frac{6r^{2}(r-1)^{2}e^{2}\vartheta^{2}\mathfrak{B}^{2}}{\hbar^{2}}-\cdots\right],\\
&=r\mathfrak{B}x+\frac{3r^{2}(1-r)e\vartheta\mathfrak{B}^{2}x}{\hbar}+\mathcal{O}(\vartheta^{2}),\\
&=\mathbf{A}_{y}+\frac{2e(r-1)}{\hbar}\theta^{yx}\mathbf{A}_{y}\partial_{x}\mathbf{A}_{y}+\frac{er}{\hbar}\theta^{yx}\mathbf{A}_{y}\partial_{y}\mathbf{A}_{x}+\mathcal{O}(\vartheta^{2}),\\
&=\mathbf{A}_{y}-\frac{e}{\hbar}\theta^{yx}\mathbf{A}_{y}\left[(2-3r)\partial_{x}\mathbf{A}_{y}+rF_{xy}\right]+\mathcal{O}(\vartheta^{2}).
\end{aligned}
\end{equation}
\endgroup

It now remains to compare the noncommutative Seiberg-Witten field strength with its commutative counterpart. The noncommutative Seiberg-Witten field strength $F^{\hbox{\tiny{nc}}}_{xy}$ is given by
\begin{equation}\label{equation:nc-S-W-field-strength}
F^{\hbox{\tiny{nc}}}_{xy}=\partial_{x}\mathbf{A}^{\hbox{\tiny{nc}}}_{y}-\partial_{y}\mathbf{A}^{\hbox{\tiny{nc}}}_{x}-\frac{ie}{\hbar}[\mathbf{A}^{\hbox{\tiny{nc}}}_{x}\!\!\overset{\;\;*^{r}}{,}\mathbf{A}^{\hbox{\tiny{nc}}}_{y}]=B\mathbb{I},
\end{equation}
with $\mathbb{I}$ being the real valued constant map in $C^{\infty}(\mathbb{R}^{2})$ that maps everything to $1$. Here, we used (\ref{equation:non-abelian-yang-mills-star-version}) to arrive at (\ref{equation:nc-S-W-field-strength}).

Now, using (\ref{equation:agreement-with-Delduc}) in (\ref{equation:nc-S-W-field-strength}), one obtains
\begin{equation}\label{equation:derivation-rel-com-noncom}
\begin{aligned}[b]
F^{\hbox{\tiny{nc}}}_{xy}&=\bar{B}\left[1-\frac{r(r-1)e\vartheta\bar{B}}{\hbar}\right]\mathbb{I},\\
&=\frac{\hbar}{2r(1-r)e\vartheta}\left[\frac{1}{\sqrt{1+\frac{4r(r-1)e\vartheta\mathfrak{B}}{\hbar}}}-1\right]\left[1+\frac{1}{2}\left\{\frac{1}{\sqrt{1+\frac{4r(r-1)e\vartheta\mathfrak{B}}{\hbar}}}-1\right\}\right]\mathbb{I},
\end{aligned}
\end{equation}
upon using (\ref{equation:def-commut-Seib-Witt-mag-field}). Let 
\begin{equation}\label{equation:substitution}
\frac{4r(r-1)e\vartheta\mathfrak{B}}{\hbar}=t
\end{equation}
in (\ref{equation:derivation-rel-com-noncom}) and simplify.

\begin{equation}\label{equation:simplify}
\begin{aligned}
F^{\hbox{\tiny{nc}}}_{xy}&=\frac{2\mathfrak{B}}{t}\left[1-\frac{1}{\sqrt{1+t}}\right]\left[1+\frac{1}{2}\left(\frac{1}{\sqrt{1+t}}-1\right)\right]\mathbb{I},\\
&=\frac{\mathfrak{B}}{t}\left[1-\frac{1}{\sqrt{1+t}}\right]\left[1+\frac{1}{\sqrt{1+t}}\right]\mathbb{I},\\
&=\left(\frac{\mathfrak{B}}{1+t}\right)\mathbb{I}.\\
\end{aligned}
\end{equation}
Substituting back the value of $t$ from (\ref{equation:substitution}) in (\ref{equation:simplify}), one finally obtains
\begin{equation}\label{equation:S-W-map-field-strength}
\begin{aligned}[b]
F^{\hbox{\tiny{nc}}}_{xy}&=\mathfrak{B}\left[1-\frac{4r(1-r)e\vartheta\mathfrak{B}}{\hbar}\right]^{-1}\mathbb{I},\\
&=\mathfrak{B}\mathbb{I}+\frac{4r(1-r)e\vartheta\mathfrak{B}^{2}}{\hbar}\mathbb{I}+\mathcal{O}(\vartheta^{2}),\\
&=F_{xy}+\frac{4er(1-r)}{\hbar}\theta^{yx}F_{xy}F_{yx}+\mathcal{O}(\vartheta^{2}).
\end{aligned}
\end{equation}
Here, $F_{xy}=-F_{yx}$ are as given in (\ref{equation:entries-commut-S-W-field-strength}). Equations (\ref{equation:first-comp-gauge-field}), (\ref{equation:y-component-S-W-map-gauge-field}) and (\ref{equation:S-W-map-field-strength}) tell us how noncommutative U(1) gauge field and Field strength are related with their commutative counterparts in the first order of the deformation parameter $\vartheta$ in a chosen gauge parametrized by a real value of $r$. These equations take the following familiar forms in the symmetric gauge configuration corresponding to $r=\frac{1}{2}$:
\begin{equation}\label{equation:S-W-map-summary}
\begin{split}
\mathbf{A}^{\hbox{\tiny{nc}}}_{x}&=\mathbf{A}_{x}-\frac{e}{2\hbar}\theta^{xy}\mathbf{A}_{x}(\partial_{y}\mathbf{A}_{x}+F_{yx})+\mathcal{O}(\vartheta^{2}),\\
\mathbf{A}^{\hbox{\tiny{nc}}}_{y}&=\mathbf{A}_{y}-\frac{e}{2\hbar}\theta^{yx}\mathbf{A}_{y}(\partial_{x}\mathbf{A}_{y}+F_{xy})+\mathcal{O}(\vartheta^{2}),\\
F^{\hbox{\tiny{nc}}}_{xy}&=F_{xy}+\frac{e}{\hbar}\theta^{yx}F_{xy}F_{yx}+\mathcal{O}(\vartheta^{2}).
\end{split}
\end{equation}
We call the 1-parameter family of maps given by (\ref{equation:y-component-S-W-map-gauge-field}), (\ref{equation:first-comp-gauge-field}) and (\ref{equation:S-W-map-field-strength}), the \textit{Seiberg-Witten maps}.

\section{Conclusion and Future Perspectives}\label{sec:conclusion}
In this paper, we have undertaken noncommutative U(1) gauge theoretic study of a spinless point particle placed in a noncommutative plane  minimally coupled to an external uniform magnetic field. In the process, we have seen how the gauge parameter $r$ gives rise to $c$-equivalent star products $*^{r}$ in the sense of \cite{Bayen-Flato1}. We also have obtained an induced family of invertible (parametrized by the same gauge parameter $r$) maps that relate the noncommutative U(1) gauge field and the corresponding field strength with their commutative counterparts. These invertible maps for $r=\frac{1}{2}$ agrees with the so-called Seiberg-Witten map constructed in \cite{Seiberg-Witten}. 

Geometrically, one has a 1-dimensional $\mathcal{A}$-module $\mathbb{M}=\mathbb{C}\times\mathcal{A}$ where the noncommutative $*$-algebra is $(C^{\infty}(\mathbb{R}^{2}),*^{r})$. Connections on $\mathbb{M}$ are then maps $\nabla:\hbox{Der}(\mathcal{A})\times\mathbb{M}\rightarrow\mathbb{M}$ satisfying some nice properties (see appendix of \cite{Kupriyanov}). There is a more interesting construction of deformation of principal G bundle $\hbox{pr}:\mathcal{P}\circlearrowleft G\rightarrow\mathcal{M}$ \cite{Waldmann1, Waldmann2} where $\mathcal{M}$ is a smooth manifold. According to the construction presented in \cite{Waldmann3}, if one has a $\star$ product defined on $\mathcal{M}$, i.e., $\star:C^{\infty}(\mathcal{M})[[\lambda]]\times C^{\infty}(\mathcal{M})[[\lambda]]\rightarrow C^{\infty}(\mathcal{M})[[\lambda]]$ with $\lambda$ being the deformation parameter, then one has an induced right $\star$-module structure $\bullet$ on $C^{\infty}(\mathcal{P})[[\lambda]]$ by
\begin{equation}\label{equation:induced-star-product}
F\bullet f=F\hbox{pr}^{*}f+\sum_{n=1}^{\infty}\lambda^{r}\varrho_{n}(F,f),
\end{equation}
with $\varrho_{n}:C^{\infty}(\mathcal{P})\times C^{\infty}(M)\rightarrow C^{\infty}(\mathcal{P})$ is a bidifferential operator along the map $\hbox{pr}$ for all $r\geq 1$ such that one also has the $G$-equivariance
\begin{equation}\label{equation:G-equivariance-condition}
g^{*}(F\bullet f)=g^{*}F\bullet f.
\end{equation}
Here, $F\in C^{\infty}(\mathcal{P})$, $f\in C^{\infty}(\mathcal{M})$ and $g\in G$. Also, $\hbox{pr}^{*}:C^{\infty}(\mathcal{M})\rightarrow C^{\infty}(\mathcal{P})$ and the right action of the group $G$ on $\mathcal{P}$ induces a map $g^{*}:C^{\infty}(\mathcal{P})\rightarrow C^{\infty}(\mathcal{P})$. 

We look forward to studying this construction in the present context where, $\mathcal{M}=\mathbb{R}^{2}$, $\lambda=\vartheta$, $G=U(1)$ and $\star=*^{r}$. We would like to define an equivalence class of $*^{r}$-module structures $\bullet^{r}$ in the sense of \cite{Waldmann3} and study its resulting algebraic properties in a future publication.

Furthermore, the Landau problem is formulated for the particle moving in a two-dimensional field. However, it can be extended to the higher dimensional space as shown, for example, in \cite{Zhang:2001xs, Elvang:2002jh}. Besides, the magnetic field can be non-uniform, as pointed out in \cite{Delduc}. It will be interesting to explore the noncommutative generalization of the higher dimensional case and the presence of a non-uniform magnetic field which we have set for a future publication.

\subsubsection*{Acknowledgement}
We acknowledge fruitful discussions with Tibra Ali that helped us understand the simple underlying gauge theoretic structures. TAC would also like to thank the High Energy Physics Theory group of the Department of Physics and Astronomy at the University of Kansas for the hospitality and support. SHHC gratefully acknowledges his six year old daughter Nefertiti's graciousness in allowing him to use her notebook to do important calculations. 

\section{Appendix}

Here we present the derivation of the expression (see \ref{equation:rel-com-nc-gauge-func}) of the {\em noncommutative gauge function} $\lambda^{\hbox{\tiny{nc}}}$ in detail. Using (\ref{equation:gauge-field-function}), one obtains
\begin{equation}\label{equation:infinitesimal-gauge-field-var-x}
\begin{aligned}
\delta \mathbf{A}^{\hbox{\tiny{nc}}}_{x}&=\mathbf{A}^{\hbox{\tiny{nc}}\prime}_{x}-\mathbf{A}^{\hbox{\tiny{nc}}}_{x}\\
&=\frac{-2(1-r^{\prime})\hbar By}{\hbar+\sqrt{\hbar^{2}-4r^{\prime}(r^{\prime}-1)e\hbar\vartheta B}}+\frac{2(1-r)\hbar By}{\hbar+\sqrt{\hbar^{2}-4r(r-1)e\hbar\vartheta B}},\\
&=-2(1-r^{\prime})By\left[1+\sqrt{1-\frac{4r^{\prime}(r^{\prime}-1)e\vartheta B}{\hbar}}\right]^{-1}+2(1-r)By\left[1+\sqrt{1-\frac{4r(r-1)e\vartheta B}{\hbar}}\right]^{-1},\\
&=\frac{\hbar}{2r^{\prime}e\vartheta}\left[1-\sqrt{1-\frac{4r^{\prime}(r^{\prime}-1)e\vartheta B}{\hbar}}\right]-\frac{\hbar}{2re\vartheta}\left[1-\sqrt{1-\frac{4r(r-1)e\vartheta B}{\hbar}}\right],\\
&=\frac{\hbar}{2r^{\prime}e\vartheta}\left[1-\left\{1-\frac{2r^{\prime}(r^{\prime}-1)e\vartheta B}{\hbar}-\frac{2r^{\prime 2}(r^{\prime}-1)^{2}e^{2}\vartheta^{2}B^{2}}{\hbar^{2}}\right\}\right]y,\\
&\hspace{.1in}-\frac{\hbar}{2re\vartheta}\left[1-\left\{1-\frac{2r(r-1)e\vartheta B}{\hbar}-\frac{2r^{2}(r-1)^{2}e^{2}\vartheta^{2}B^{2}}{\hbar^{2}}\right\}\right]y+\mathcal{O}(\vartheta^{2}),\\
&=(r^{\prime}-1)By+r^{\prime}(r^{\prime}-1)^{2}\frac{e\vartheta B^{2}y}{\hbar}-(r-1)By-\frac{r(r-1)^{2}e\vartheta B^{2}y}{\hbar}+\mathcal{O}(\vartheta^{2}),\\
&=(r^{\prime}-r)By+\frac{e\vartheta B^{2}y}{\hbar}[r^{\prime}(r^{\prime}-1)^{2}-r(r-1)^{2}]+\mathcal{O}(\vartheta^{2}).
\end{aligned}
\end{equation}
Similarly, using (\ref{equation:gauge-field-function}), one can proceed to obtain the expression for the infinitesimal variation of the $y$-component of the noncommutative gauge field:
\begin{equation}\label{equation:infinitesimal-gauge-field-variant-y}
\begin{aligned}
\delta\mathbf{A}^{\hbox{\tiny{nc}}}_{y}=(r^{\prime}-r)Bx+\frac{e\vartheta B^{2}x}{\hbar}[r^{\prime 2}(r^{\prime}-1)-r^{2}(r-1)]+\mathcal{O}(\vartheta^{2}).
\end{aligned}
\end{equation}
One, now takes the infinitesimal variation $\delta \mathbf{A}^{\hbox{\tiny{nc}}}_{x}$ of the $x$-component of the noncommutative gauge field from (\ref{equation:infinitesimal-gauge-field-var-x}) and expand it in $\epsilon=r^{\prime}-r$:
\begin{equation}\label{equation:infinitesimal-gauge-x-derivation}
\begin{aligned}
\delta\mathbf{A}^{\hbox{\tiny{nc}}}_{x}&=By\epsilon+\frac{B^{2}e\vartheta y}{\hbar}[(r+\epsilon-1)^{2}(r+\epsilon)-(r-1)^{2}r]+\mathcal{O}(\vartheta^{2}),\\
&=By\epsilon+\frac{B^{2}e\vartheta y}{\hbar}[r(r-1)^{2}+r\epsilon^{2}+2r(r-1)\epsilon+\epsilon(r-1)^{2}+\epsilon^{3}+2(r-1)\epsilon^{2}-(r-1)^{2}r]+\mathcal{O}(\vartheta^{2}),\\
&=By\epsilon+\frac{B^{2}e\vartheta y}{\hbar}(r\epsilon^{2}+2r^{2}\epsilon-2r\epsilon+\epsilon r^{2}-2r\epsilon+\epsilon+\epsilon^{3}+2r\epsilon^{2}-2\epsilon^{2})+\mathcal{O}(\vartheta^{2}),\\
&=By\epsilon+\frac{B^{2}ye\vartheta\epsilon}{\hbar}(3r^{2}-4r+1)+\mathcal{O}(\epsilon^{2})+\mathcal{O}(\vartheta^{2}).
\end{aligned}
\end{equation}
Since $\epsilon=r^{\prime}-r$ is an infinitesimal, $\mathcal{O}(\epsilon^{2})$ can be neglected and (\ref{equation:infinitesimal-gauge-x-derivation}) reduces to
\begin{equation}\label{equation:infinitesimal-x-form}
\delta\mathbf{A}^{\hbox{\tiny{nc}}}_{x}=By\epsilon+\frac{B^{2}ye\vartheta\epsilon}{\hbar}(3r^{2}-4r+1)+\mathcal{O}(\vartheta^{2}).
\end{equation}

Similarly, for the y-component of the noncommutative gauge field, (\ref{equation:infinitesimal-gauge-field-variant-y}) can be expanded in $\epsilon=r^{\prime}-r$. Here, again one retains only the terms linear in the infinitesimal $\epsilon$ to obtain,
\begin{equation}\label{equation:infinitesimal-y-form}
\delta\mathbf{A}^{\hbox{\tiny{nc}}}_{y}=Bx\epsilon+\frac{B^{2}xe\vartheta\epsilon}{\hbar}(3r^{2}-2r)+\mathcal{O}(\vartheta^{2}).
\end{equation}

But according to (\ref{equation:def-nc-gauge-function}), the infinitesimal variation of the $x$ component of the noncommutative gauge field is given by
\begin{equation}\label{equation:x-comp-gauge-var-def}
\delta\mathbf{A}^{\hbox{\tiny{nc}}}_{x}=\partial_{x}\lambda^{\hbox{\tiny{nc}}}+\frac{ie}{\hbar}[\lambda^{\hbox{\tiny{nc}}}\!\overset{\;\;*^{r}}{,}\mathbf{A}^{\hbox{\tiny{nc}}}_{x}].
\end{equation} 

Now, write 
\begin{equation}\label{equation:nc-gauge-func-def}
\lambda^{\hbox{\tiny{nc}}}=\widetilde{B}\epsilon xy,
\end{equation} 

and substitute in (\ref{equation:x-comp-gauge-var-def}). Use (\ref{equation:gauge-field-function}) and the definition of the 1-parameter family of star-products (\ref{equation:star-prod-gen-def}) together with (\ref{equation:nc-gauge-func-def}) in (\ref{equation:x-comp-gauge-var-def}) to obtain
\begin{equation}\label{equation:x-com-gauge-var-right-side}
\begin{aligned}
\delta\mathbf{A}^{\hbox{\tiny{nc}}}_{x}&=\widetilde{B}\epsilon y+\frac{ie}{\hbar}\left[\widetilde{B}\epsilon xy\!\overset{\;\;*^{r}}{,}\frac{2(r-1)\hbar By}{\hbar+\sqrt{\hbar^{2}-4r(r-1)e\hbar\vartheta B}}\right],\\
&=\widetilde{B}\epsilon xy+\frac{ie\vartheta}{\hbar}\times i\epsilon(r-1)\widetilde{B}\times\frac{2\hbar By}{\hbar+\sqrt{\hbar^{2}-4r(r-1)e\hbar\vartheta B}},\\
&=\widetilde{B}\left[1+\frac{2e(1-r)B\vartheta}{\hbar+\sqrt{\hbar^{2}-4r(r-1)e\hbar\vartheta B}}\right]\epsilon y.\\
\end{aligned}
\end{equation}

Now comparing (\ref{equation:x-com-gauge-var-right-side}) with (\ref{equation:infinitesimal-x-form}), one obtains,
\begin{equation}\label{equation:x-comp-cal}
\begin{aligned}
\widetilde{B}\left[1+\frac{2e(1-r)B\vartheta}{\hbar+\sqrt{\hbar^{2}-4r(r-1)e\hbar\vartheta B}}\right]=B+\frac{eB^{2}\vartheta}{\hbar}(r-1)(3r-1)+\mathcal{O}(\vartheta^{2}).
\end{aligned}
\end{equation}
Rearrangement of the terms in (\ref{equation:x-comp-cal}) leads to
\begin{equation}\label{equation:expr-tilde-B}
\begin{aligned}
\widetilde{B}&=\left[B+\frac{eB^{2}\vartheta}{\hbar}(r-1)(3r-1)\right]\left[1+\frac{2e(1-r)B\vartheta}{\hbar+\sqrt{\hbar^{2}-4r(r-1)e\hbar\vartheta B}}\right]^{-1},\\
&=\left[B+\frac{eB^{2}\vartheta}{\hbar}(r-1)(3r-1)\right]\left[1+\frac{2e(1-r)B\vartheta}{\hbar}\left\{1+\sqrt{1-\frac{4r(r-1)e\vartheta B}{\hbar}}\right\}^{-1}\right]^{-1},\\
&=\left[B+\frac{eB^{2}\vartheta}{\hbar}(r-1)(3r-1)\right]\left[1+\frac{2e(1-r)B\vartheta}{\hbar}\left\{\frac{1}{2}+\frac{4er(r-1)\vartheta B}{8\hbar}+...\right\}\right]^{-1},\\
&=B-e(1-r)\frac{B^{2}\vartheta}{\hbar}+\frac{eB^{2}\vartheta}{\hbar}(r-1)(3r-1)+\mathcal{O}(\vartheta^{2}),\\
&=B+3er(r-1)\frac{B^{2}\vartheta}{\hbar}+\mathcal{O}(\vartheta^{2}),
\end{aligned}
\end{equation}
so that one finally obtains the following expression for the noncommutative gauge function $\lambda^{\hbox{\tiny{nc}}}$:
\begin{equation}\label{equation:final-exprs-nc-gauge-fun}
\lambda^{\hbox{\tiny{nc}}}=B\epsilon xy+3e\epsilon r(r-1)\vartheta\frac{B^{2}xy}{\hbar}+\mathcal{O}(\vartheta^{2}).
\end{equation}
We leave it to the reader to verify that the above expression (\ref{equation:final-exprs-nc-gauge-fun}) of noncommutative gauge function indeed satisfies the defining equation of the infinitesimal variation of the y-component of the noncommutative gauge field (see \ref{equation:def-nc-gauge-function}):
 \begin{equation}\label{equation:y-comp-gauge-var-def}
\delta\mathbf{A}^{\hbox{\tiny{nc}}}_{y}=\partial_{y}\lambda^{\hbox{\tiny{nc}}}+\frac{ie}{\hbar}[\lambda^{\hbox{\tiny{nc}}}\!\overset{\;\;*^{r}}{,}\mathbf{A}^{\hbox{\tiny{nc}}}_{y}],
\end{equation} 
with $\mathbf{A}^{\hbox{\tiny{nc}}}_{y}$ is as in (\ref{equation:gauge-field-function}) and $\delta\mathbf{A}^{\hbox{\tiny{nc}}}_{y}$ is as in (\ref{equation:infinitesimal-y-form}).

\end{document}